\documentclass[twocolumn,prc,preprintnumbers,amssymb,
               floatfix]{revtex4}
\usepackage{dcolumn}
\usepackage{bm}
\usepackage{mathrsfs}
\usepackage{graphicx,epsfig,latexsym,amssymb}
\usepackage{multirow,amsmath,array,booktabs,color}
\usepackage[section]{placeins}
\usepackage{hyperref}
\usepackage{marvosym}
\usepackage{color}
\usepackage{epstopdf}
\usepackage{ulem}

\begin{document}
	
	\preprint{ }
	
	\title{Zero-sound modes for the nuclear equation of state at
		supra-normal densities}
	% Force line breaks with \\

	\author{Jing Ye$^1$, J. Margueron$^2$, Niu Li$^{1}$, and W. Z. Jiang$^1$\footnote{ wzjiang@seu.edu.cn}}
	\affiliation{$^1$ Department of Physics, Southeast University, Nanjing 211189, China\\
$^2$ Institut de Physique Nucl\'eaire de Lyon, CNRS/IN2P3, Universit\'e de Lyon, \\ Université Claude Bernard Lyon 1, F-69622 Villeurbanne Cedex, France}

	\date{\today}% It is always \today, today,
	
	\begin{abstract}
	The meaningful correlations between the zero-sound modes and the stiffness of the nuclear equation of state (EOS) are uncovered in nuclear matter with the relativistic mean-field theory. It is demonstrated that the high-density zero-sound modes merely exist in models with the stiff EOS. While  the stiff EOS can be softened by including $\omega$-meson self-interactions (the $\omega^4$ term), the weakened coupling of the $\omega$-meson self-interactions reignites the zero sound at high density.  These  results suggest that the high-density zero-sound modes can be used to probe the stiffness of the EOS at supranormal densities. The implications and effects of zero sounds are also discussed in heavy ion collisions and neutron stars.
	\end{abstract}

	\maketitle

\section{\label{sec:level1}INTRODUCTION}
	A central issue in the theory of astrophysical compact objects and heavy-ion reactions is the equation of state (EOS) of asymmetric matter which is crucial for the description of the exotic nuclei off $\beta$ stability, the dynamical evolution of many violent astrophysical events, and the structure of the emerging compact stars~\cite{horowitz2001neutron,jiang2010superheavy, wang2015drip,shen2011second,lattimer2012nuclear,oertel2017equations}. After decades of effort, appreciable progress has been achieved on constraining the EOS near or beneath the saturation density based on astrophysical observations or terrestrial laboratory experiments~\cite{hartnack2006hadronic,danielewicz2002determination,lattimer2014neutron}. However, accurate extraction of the high-density EOS through the energetic heavy-ion collisions in the terrestrial laboratories or the observations of neutron star properties is rather challenging~\cite{li2008recent,danielewicz2002determination,steiner2010equation}, although the multimessenger observations by virtue of the currently operating satellites (NICER~\cite{riley2021nicer,miller2019psr}) and the gravitational-wave (GW) laser interferometers (advanced LIGO and Virgo~\cite{abbott2019properties,abbott2018gw170817,abbott2017gw170817}) may hopefully impose new constraints on the structure of neutron stars and the EOS, for instance, see Ref.~\cite{Perego2021}. Due to the high nonlinearity of the nuclear many-body problems, the theoretical extrapolation of the EOS to the high-density region seems to be quite diversified, depending upon the models or approaches used. In this case, it is very meaningful to search for new theoretical constraints and sensitive probes to the high-density EOS that can possibly be verified by experiments.
	
	 The electric multipolar excitation of the atomic nucleus reflects or constrains properties of the EOS ~\cite{roca2018nuclear}, such as the nuclear matter compression modulus~\cite{stone2014incompressibility,garg2018compression}, the symmetry energy~\cite{trippa2008giant,lattimer2013constraining}, and the effective mass of the nucleon~\cite{harakeh2001giant,roca2013giant} which are correlated tightly with  the isoscalar  giant multipole resonances,  the isovector  giant dipole resonance, and the isoscalar giant quadrupole resonance, respectively. Currently, the widely studied nuclear collective modes are  mainly limited to finite nuclear systems near or below saturation density. The constraints on the EOS from these collective modes just apply  to relevant density domains. In a search of the constraints on the high-density EOS,  our attention will be paid to the zero sound at suprasaturation densities, as it is pervasive in the meson-nucleon interacting system~\cite{chin1977relativistic,lim1989collective}.  As a digression, it is interesting to mention the zero sound in condensed matter. At low temperatures, the usual propagation of ordinary sound waves in a gas is suppressed due to the rarity of collisions. However, in a Fermi gas, another type of oscillation mode: zero sound,  predicted by Landau early in 1957~\cite{landau1957oscillations} and subsequently observed in liquid $^3$He~\cite{roach1976observation,abel1966propagation}, can still propagate even close to zero temperature. The concept of zero sound has been applied in the past to isotropic two-component Fermi liquids~\cite{dobbs2000helium}, holographic quantum liquids~\cite{karch2009holographic}, and nuclear matter~\cite{chin1977relativistic}. In particular, it was reported that zero-sound modes in neutron stars can  influence the neutrino irradiation and heat conduction in the star cooling process in the presence of the nucleon superfluids~\cite{bedaque2003goldstone,Aguil2009prl,leinson2011zero}. With increasing attention on the sound velocity in neutron stars~\cite{tews2018constraining,bedaque2015sound},  the study of the zero sounds in dense matter is of special interest.
	
	 In the past, various approaches from the macroscopic to microscopic models have been developed to study collective modes. The hydrodynamic approaches are usual macroscopic models where the collective modes are obtained by solving the master equation which may be regarded as a semiclassic form of the time-dependent Hartree-Fock (HF)  theory~\cite{myers1977droplet,hofmann1979theor,di1985semi,yanhuang1989semi,kolomietz2001sound}. Combining rather conveniently with the transport models, the macroscopic approaches can still renew the application in  reproducing the data of collective modes~\cite{papa2005pre,baran2009dyn,tian2010dyn,he2014prl}. Microscopically, the collective modes are mainly derived from the time-dependent HF approach, the  random phase approximation (RPA), and self-consistent extensions with more complicated configurations~\cite{bracco2019iso}. Strikingly, the relativistic RPA is equivalent to  the time-dependent relativistic mean-field (RMF) theory in the limit of small amplitude oscillations~\cite{ring2001time,vret2005rel}.
The nonrelativistic and relativistic RPA have been used very successfully in studies of various collective modes in finite nuclei~\cite{cavinato1982photoreactions,de1998relativistic, ma2001isoscalar,vretenar2002toroidal,vretenar2003spin,liang2008spin}. In this work, we employ the relativistic RPA  based on the RMF theory to explore the characteristics of zero-sound modes in nuclear matter especially at suprasaturation densities. The emphasis will be placed on the relationship between the stiffness of the EOS at suprasaturation densities and the zero sound, while the  variation of the stiffness of the high-density EOS can be  simulated  by mainly adjusting the vector potential through the self-interaction of the vector meson~\cite{yang2017novel}.
	
	The remainder of the paper is organized as follows. In Sec.~\ref{sec:level2}, we will introduce briefly the formalism of the relativistic mean-field models and the random phase approximation. In Sec.~\ref{sec:level3}, numerical results and discussions are presented. At last, a summary is given in Sec.~\ref{sec:level4}.
	
	\section{\label{sec:level2}FORMALISM}
	\subsection{\label{sec:level2-1}Effective model of strong interaction}
	In the  RMF approach, the isoscalar\mbox{-}scalar $\sigma$, the isoscalar\mbox{-}vector $\omega$, and the isovector\mbox{-}vector $\rho$ ($b_{0}$) mesons mediate the nuclear interactions to quantitatively  describe nuclear matter and finite nuclei. The effective Lagrangian can be written as~\cite{horowitz2001neutron,serot1992relativistic,bodmer1991relativistic,jiang2005charge}
	\begin{eqnarray}\label{eq-L}
		\mathcal{L}=&& \bar{\psi}\left[i \gamma_{\mu} \partial^{\mu}-M+g_{\sigma} \sigma-g_{\omega} \gamma_{\mu} \omega^{\mu}-g_{\rho} \gamma_{\mu} \tau_{3} b_{0}^{\mu}\right] \psi \nonumber\\
		&&+\frac{1}{2}\left(\partial_{\mu} \sigma \partial^{\mu} \sigma-m_{\sigma}^{2} \sigma^{2}\right)-\frac{1}{3} g_{2} \sigma^{3}-\frac{1}{4} g_{3} \sigma^{4}\nonumber\\
		&&-\frac{1}{4} F_{\mu \nu} F^{\mu \nu}+\frac{1}{2} m_{\omega}^{2} \omega_{\mu} \omega^{\mu}+\frac{1}{4} c_{3}\left(\omega_{\mu} \omega^{\mu}\right)^{2}\nonumber\\
		&&-\frac{1}{4} B_{\mu \nu} B^{\mu \nu}+\frac{1}{2} m_{\rho}^{2} b_{0 \mu} b_{0}^{\mu}\nonumber\\
		&& +4 \Lambda_{V} g_{\rho}^{2} g_{\omega}^{2} \omega_{\mu} \omega^{\mu} b_{0 \nu} b_{0}^{\nu},
	\end{eqnarray}
	where nucleons and mesons ($\psi,\sigma ,\omega,b_{0}$) have their free masses $M$, $m_{\sigma}$, $m_{\omega}$ and $m_{\rho}$, respectively.
  $F_{\mu \nu}$ and $B_{\mu \nu}$ are the respective strength tensors of the vector mesons $\omega$ and $\rho$,
	\begin{equation}
		F_{\mu \nu}=\partial_{\mu} \omega_{\nu}-\partial_{\nu} \omega_{\mu}, \quad B_{\mu \nu}=\partial_{\mu} b_{0 v}-\partial_{\nu} b_{0 \mu}.
	\end{equation}
		
	The nonlinear equations of motion for the nucleon and mesons are deduced from the  standard Euler-Lagrange formula, and in the mean-field approximation, meson fields are replaced by their expectation values.  With these mean-field quantities solved by iteration, the resulting effective meson masses $m_{i}^{*}$ are given by the formulas
	\begin{subequations}
		\begin{align}
			&m_{\sigma}^{* 2}=m_{\sigma}^{2}+2 g_{2} \sigma_{0}+3 g_{3} \sigma_{0}^{2}, \label{eq:ms}\\
			&m_{\omega}^{* 2}=m_{\omega}^{2}+3 c_{3} \omega_{0}^{2}+8 \Lambda_{V} g_{\omega}^{2} g_{\varrho}^{2} b_{0}^{2} ,\\
			&m_{\varrho}^{* 2}=m_{\varrho}^{2}+8 \Lambda_{V} g_{\omega}^{2} g_{\varrho}^{2} \omega_{0}^{2}.
		\end{align}
	\end{subequations}
The energy density $\varepsilon$ and pressure $P$ of nuclear matter are written as	
	\begin{eqnarray}\label{eq-E}
		\varepsilon=&& \sum_{i=p, n} \frac{2}{(2 \pi)^{3}} \int^{k_{F_{i}}} d^{3} k E_{i}^{*}+\frac{1}{2} m_{\omega}^{2} \omega_{0}^{2}+\frac{1}{2} m_{\sigma}^{2} \sigma_{0}^{2} \nonumber\\
		&&+\frac{1}{2} m_{\rho}^{2} b_{0}^{2}+\frac{1}{3} g_{2} \sigma_{0}^{3}+\frac{1}{4} g_{3} \sigma_{0}^{4}+\frac{3}{4} c_{3} \omega_{0}^{4}\nonumber\\
		&&+12 \Lambda_{V} g_{\rho}^{2} g_{\omega}^{2} \omega_{0}^{2} b_{0}^{2},
	\end{eqnarray}
	
	\begin{eqnarray}\label{eq-P}
		P=&& \frac{1}{3} \sum_{i=p, n} \frac{2}{(2 \pi)^{3}} \int^{k_{F_{i}}} d^{3} k \frac{\mathbf{k}^{2}}{E_{i}^{*}}+\frac{1}{2} m_{\omega}^{2} \omega_{0}^{2}-\frac{1}{2} m_{\sigma}^{2} \sigma_{0}^{2}\nonumber\\
		&&+\frac{1}{2} m_{\rho}^{2} b_{0}^{2}-\frac{1}{3} g_{2} \sigma^{3}-\frac{1}{4} g_{3} \sigma^{4}+\frac{1}{4} c_{3} \omega_{0}^{4}\nonumber \\
		&&+4 \Lambda_{V} g_{\rho}^{2} g_{\omega}^{2} \omega_{0}^{2} b_{0}^{2}
	\end{eqnarray}
	with  $E_{i}^{*}=\sqrt{\mathbf{k}^{2}+\left(M_{i}^{*}\right)^{2}}.$ For neutron star matter, Eqs.~(\ref{eq-E}) and (\ref{eq-P}) can be easily extended to include leptons based on the chemical and $\beta$ equilibriums~\cite{zhang2011prc,xiang2014prc}.
	
	\subsection{\label{sec:level2-2}Polarizations in RPA}
	In the relativistic RPA approach, one can, in principle, involve the particle-hole and particle-antiparticle excitations of the fermions by the polarization functions, while with the framework of the RMF models  the particle-antiparticle excitations are usually ignored from the Dirac sea.   The interacting polarization is determined through the Dyson equation, and, for instance, the longitudinal polarization that is usually used to search for the collective modes is given as
	\begin{equation}
		\tilde{\Pi}_{L}={\Pi}_{L}+{\Pi}_{L}D_{L}\tilde{\Pi}_{L},
	\end{equation}
	where the polarization $\Pi$ and  propagator $D$ are in the matrix form.  In generalized matter including electrons,  the lowest-order longitudinal polarization matrix is written as
	\begin{equation}\label{eq:pi}
		\Pi_{L}=\left(\begin{array}{cccc}
			\Pi_{00}^{e}& 0& 0& 0\\
            0& \Pi_{s}^{n}+\Pi_{s}^{p} & \Pi_{m}^{p} & \Pi_{m}^{n} \\
			0& \Pi_{m}^{p} & \Pi_{00}^{p} & 0 \\
			0& \Pi_{m}^{n} & 0 & \Pi_{00}^{n}
		\end{array}\right),
	\end{equation}
 where the individual polarization entries are given by
	\begin{subequations}
		\begin{align}
			i \Pi_{s}\left(\bar{q}, q_{0}\right) &=\int \frac{d^{4} p}{(2 \pi)^{4}} \operatorname{Tr}[G(p) G(p+q)] ,\\
			i \Pi_{m}\left(\bar{q}, q_{0}\right) &=\int \frac{d^{4} p}{(2 \pi)^{4}} \operatorname{Tr}\left[G(p) \gamma_{0} G(p+q)\right] ,\\
			i \Pi_{00}\left(\bar{q}, q_{0}\right) &=\int \frac{d^{4} p}{(2 \pi)^{4}} \operatorname{Tr}\left[G(p) \gamma_{0} G(p+q) \gamma_{0}\right]
		\end{align}
	\end{subequations}
	with Tr indicating the trace over Dirac indices. Here, the nucleon  Green function reads
	\begin{eqnarray}
		 G_{i}(k) =\left(\gamma_{\mu}k^{\mu}+M_{i}^{*}\right) \left[\frac{1}{k_{\mu}^{ 2}-M_{i}^{* 2}+i\varepsilon}\right.\nonumber\\
		\left.+\frac{i \pi}{E_{ki}^{*}} \delta\left(k_{0}-E_{ki}^{*}\right) \theta\left(k_{\mathrm{F}i}-|\boldsymbol{k}|\right)\right],i=p,n,
	\end{eqnarray}
	where $M^{*}=M-g_{\sigma}\sigma_{0}$ is the nucleon effective mass, and $k_{Fi}$ is the Fermi momentum. In nuclear matter without electrons, $\Pi_L$ in Eq.~(\ref{eq:pi}) reduces to a $3\times3$ matrix.
	
	The propagator matrix $D_L$ is in a relatively simple form if there are no crossing coupling terms between different mesons~\cite{horowitz1991neutrino}. With the  crossing coupling term of the $\rho$ and $\omega$ mesons in Eq.~(\ref{eq-L}), we first interpret the concise derivation. With the path integral method, it has been proven that the generating functional of proper vertices $\Gamma^0[\Phi]$ in the tree approximation is equal to canonical action $S[\Phi]$,
		\begin{equation}
		\Gamma^0[\Phi]=S[\Phi]=\int d^4\!x\mathcal{L}.
	    \end{equation}
In this case, the propagator $D(x,y)$ and two-point proper vertex $\Gamma^0_{2}$ are  simply the mutual reciprocal, i.e.,
 	\begin{equation}
	\Gamma^0_{2}(x,y)=D^{-1}(x,y)
	\end{equation}	
 with $\Gamma^0_{2}(x,y)=\frac{\delta^{2}\Gamma^0}{\delta\phi(x)\delta\phi(y)}$, which follows from a general relation
		\begin{equation}
		\int d^4\!z \Gamma_2(x,z)D(z-y)=\delta^4(x-y).
	    \end{equation}
 Accordingly, the  propagator $D_{L}$ with the crossing coupling term, obtained from the inverse of $\Gamma_2^0$ in the momentum representation,  is given as~\cite{carriere2003low}
	\begin{equation}\label{eq:dl}
		D_{L}=\left(\begin{array}{cccc}
           \tilde{\chi}_{\gamma}&0&-\tilde{\chi}_{\gamma}&0 \\
			0&\chi_{\sigma} & 0 & 0 \\
			-\tilde{\chi}_{\gamma}&0 & \tilde{\chi}_{\gamma}+\tilde{\chi}_{V}+2 \tilde{\chi}_{\omega \rho} & \tilde{\chi}_{I} \\
			0&0 & \tilde{\chi}_{I} & \tilde{\chi}_{V}-2 \tilde{\chi}_{\omega \rho}
		\end{array}\right)
	\end{equation}
	with $\chi_V=\chi_{\omega}+\chi_{\rho}$, $\chi_I=\chi_{\omega}-\chi_{\rho}$, and $\tilde{\chi}_{i}=\frac{q_{\mu}^{2}}{\bar{q}^{2}}\chi_{i}, i=\omega,\rho$. Expressions for the various meson propagators with nonlinear meson couplings are given as follows:
	\begin{subequations}
		\begin{align}
            &\chi_{\gamma}=\frac{e^{2}}{q_{\mu}^{2}},\\
			&\chi_{\sigma}=\frac{g_{\sigma}^{2}}{\left(q_{\mu}^{2}-m_{\sigma}^{*2}\right)}, \label{eq:Xs}\\
			&\chi_{\omega}=\frac{g_{\omega}^{2}\left(q_{\mu}^{2}-m_{\varrho}^{*2}\right)}{\left(q_{\mu}^{2}-m_{\varrho}^{*2}\right)\left(q_{\mu}^{2}-m_{\omega}^{*2}\right)-\left(16 \Lambda_{V} g_{\omega}^{2} g_{\varrho}^{2} \omega_{0} b_{0}\right)^{2}} ,\\
			&\chi_{\varrho}=\frac{g_{\varrho}^{2}\left(q_{\mu}^{2}-m_{\omega}^{*2}\right)}{\left(q_{\mu}^{2}-m_{\varrho}^{*2}\right)\left(q_{\mu}^{2}-m_{\omega}^{*2}\right)-\left(16 \Lambda_{V} g_{\omega}^{2} g_{\varrho}^{2} \omega_{0} b_{0}\right)^{2}} ,\\
			&\chi_{\omega \varrho}=\frac{-16 \Lambda_{V} g_{\omega}^{3} g_{\varrho}^{3} \omega_{0} b_{0}}{\left(q_{\mu}^{2}-m_{\varrho}^{*2}\right)\left(q_{\mu}^{2}-m_{\omega}^{*2}\right)-\left(16 \Lambda_{V} g_{\omega}^{2} g_{\varrho}^{2} \omega_{0} b_{0}\right)^{2}}.
		\end{align}
	\end{subequations}
		
	The nuclear systems undergo  transitions at  small-amplitude density fluctuations  by encountering the zeros of the following dielectric function
	\begin{equation}\label{eqdi1}
		\varepsilon_L=\operatorname{det}\left(1-D_{L} \Pi_{L}\right)=0.
	\end{equation}
As a collective mode that is determined by the zeros of the dielectric function, zero sound follows the branch of the dispersion relation that has the limit $q_0\rightarrow 0$ for  $q \rightarrow 0$, apart from the optical branch (called meson branch in the literature~\cite{lim1989collective}). If the imaginary part of the dielectric function also vanishes, the zero sound is undamped, while it is damped for the non-vanishing imaginary part. In particular, the static uniform matter ($q_0=0)$ becomes unstable at subsaturation densities when the positive $\varepsilon_L$ changes its sign~\cite{horowitz2001neutron}.  Note that the microscopic RPA calculation of the zero sound herein can alternatively be interpreted by Landau's zero sound kinetic equations~\cite{matsui1981fermi}.
	With the expressions of $D_L$ and $\Pi_L$, we can write explicitly the longitudinal dielectric function in symmetric nuclear matter  as
	\begin{eqnarray}\label{eqeL1}
		\varepsilon_L=&&\left(1-2 \Pi_{s} \chi_{s}\right) \left[1+4 \Pi_{L}^{2} \chi_{w} \chi_{\rho}+2 \Pi_{L}\left(\chi_{\omega}+\chi_{\rho}\right)\right]\nonumber\\
		&&-4\left(1+2 \chi_{\rho} \Pi_{L}\right) \chi_{s} \tilde{\chi}_{\omega} \Pi_{m}^{2},
	\end{eqnarray}
  which neglects the electron composition in Eq.(\ref{eq:pi}) and photon exchange in Eq.(\ref{eq:dl}) and is rewritten as
	\begin{eqnarray}\label{eqeL2}
		\varepsilon_L= \varepsilon_{s}\varepsilon_{v}-\varepsilon_{m},
	\end{eqnarray}
	where $\varepsilon_{s}$, $\varepsilon_{v}$, and $\varepsilon_{m}$, are the scalar, vector and scalar-vector mixed components,  respectively.

	\section{\label{sec:level3}RESULTS AND DISCUSSIONS}
\begin{figure}[htbp!]
		\includegraphics[width=0.48\textwidth]{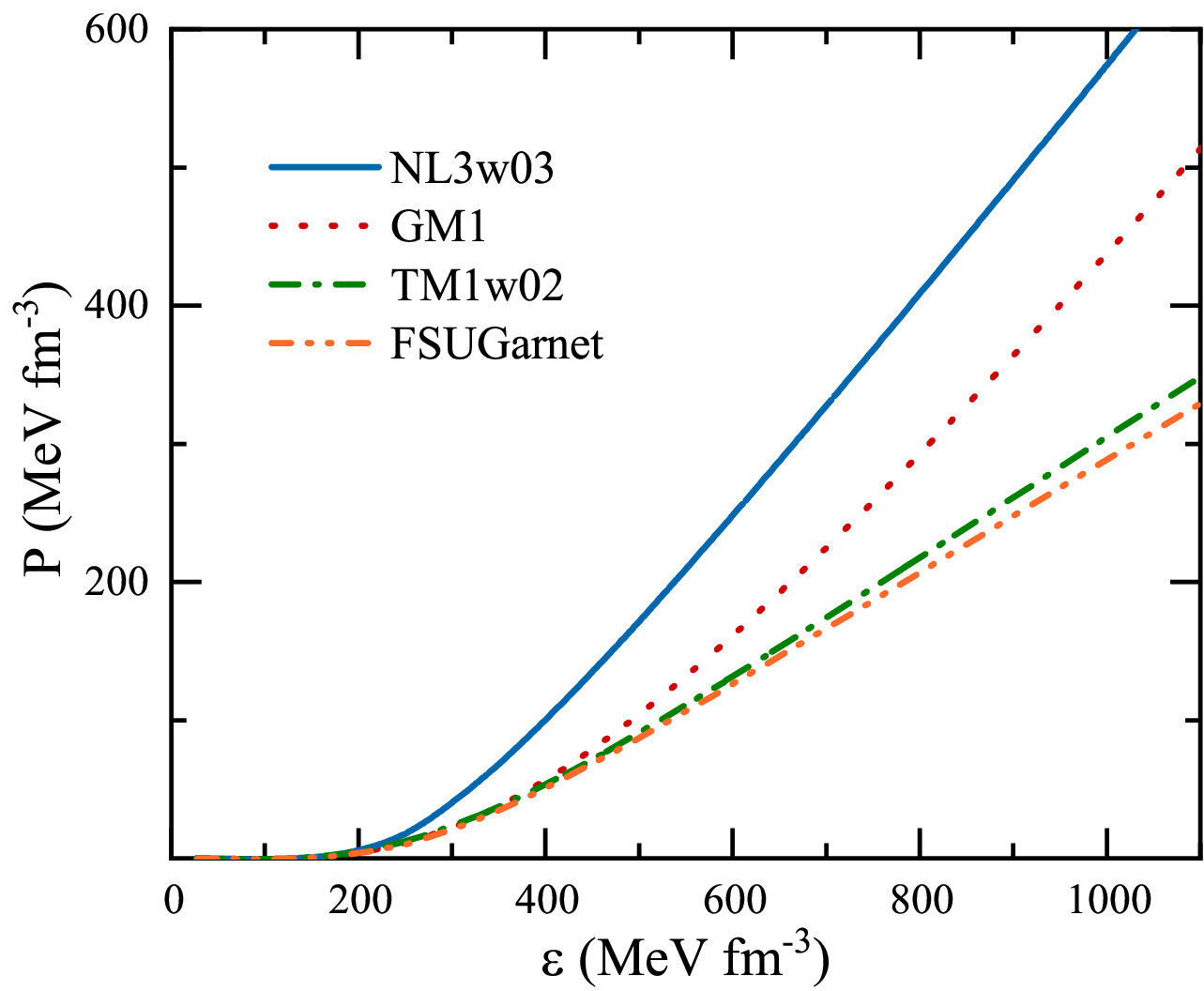}
		\centering
		\caption{\label{EOS} (Color online) The relation between the pressure and the energy  density in symmetric nuclear matter with various RMF parameter sets, NL3w03,  GM1, TM1w02, and FSUGarnet.	}
	\end{figure}

	In this work, we compare the properties of zero sound with four typical parameter sets,  NL3~\cite{lalazissis1997new}, GM1~\cite{glendenning1991reconciliation}, TM1~\cite{sugahara1994relativistic}, and FSUGarnet~\cite{chen2015searching} with a rough classification of the stiff and soft EOSs at high densities. In order to better satisfy the radius constraints of neutron stars~\cite{riley2019nicer,miller2019psr}, TM1 and NL3 are modified by introducing the isoscalar-isovector ($\omega-\rho$)  coupling term which is used to  produce the softer symmetry energy and smaller radii of neutron stars~\cite{horowitz2001radii}. For a given  $\omega-\rho$ coupling constant  $\Lambda_v$, the $\rho NN$ coupling constant $g_\rho$ is readjusted to keep the symmetry energy unchanged at $k_F $= 1.15 fm$^{-1}$, following Ref.~\cite{horowitz2001neutron}. The modified models NL3w03 and TM1w02 based respectively  on NL3 and TM1 are renamed according to the value of $\Lambda_v$, see Table~\ref{tab:Models}, where parameters and saturation properties of these parameter sets are listed. Shown in Fig.~\ref{EOS}  is the relation between the pressure and  energy density which is usually regarded as the nuclear EOS. The sound velocity square $v_s^2$ with $v_s^2=\partial P/ \partial \epsilon$ being the partial derivative of the pressure with respect to the energy density is used to describe the stiffness of the EOS. It is seen that the EOS with parameter sets TM1w02 and FSUGarnet is clearly softer than that with NL3w03 and GM1 with increasing density. The softening stems from the inclusion of the nonlinear self-interaction term of the $\omega$ meson ($\sim c_3\omega^4$) that lowers the repulsion provided by the $\omega$ meson at high densities, while the excess softening with the FSUGarnet as compared to that with the TM1w02 can be attributed dominantly to the larger parameter $c_3$ in FSUGarnet.
	
	\begin{table*}
		\caption{\label{tab:Models}Parameters and saturation properties for various parameter sets. The meson masses $m_i~(i=\sigma,\omega,\rho$), the incompressibility $K_0$,  and the symmetry energy $E_{sym}$ are  in units of MeV. The saturation density $\rho_0$ is in units of fm$^{-3}$. }
		\begin{ruledtabular}
			\begin{tabular}{ccccccccccccccc}
				
				&$g_{\sigma}$&$g_{\omega}$&$g_{\rho}$ &$m_{\sigma}$&$m_{\omega}$&$m_{\rho}$&$g_{2}$&$g_{3}$
				&$c_{3}$&$\Lambda_{V}$&$\rho_{0}$&$K_0$&$M^{*}/M$&$E_{sym}$\\ \hline

				NL3w03&$10.217 $ &$12.868 $  &$5.664 $ &$508.270 $ &$782.501 $ &$763 $ &$10.431 $ &$-28.890 $
				&- & 0.03&$0.148 $ &$272.56 $ &$0.60 $ &$31.8$\\

				GM1&$8.700$ &$10.603$  &$4.060$ &$500.000$ &$782.500$ &$763$ &$9.235$ &$-6.131$
				&- & -&$0.153$ &$300.28$ &$0.70$ &$32.5$\\		
	
				TM1w02&$10.029  $ &$12.614  $  &$5.277  $ &$511.198  $ &$783.000 $ &$770  $ &$7.233  $ &$0.618 $
				&71.31  &0.02 &$0.145  $ &$281.20  $ &$0.63 $ &$30.7$\\
				
				FSUGarnet&10.505  &13.700  &6.945 &496.939  &782.500  &763  &9.576 &$-7.207$
				&137.981  & 0.04338 &0.153  &229.63  &0.58  &30.9\\

			\end{tabular}
		\end{ruledtabular}
	\end{table*}
	
\begin{figure}[htbp!]
		\includegraphics[width=0.48\textwidth]{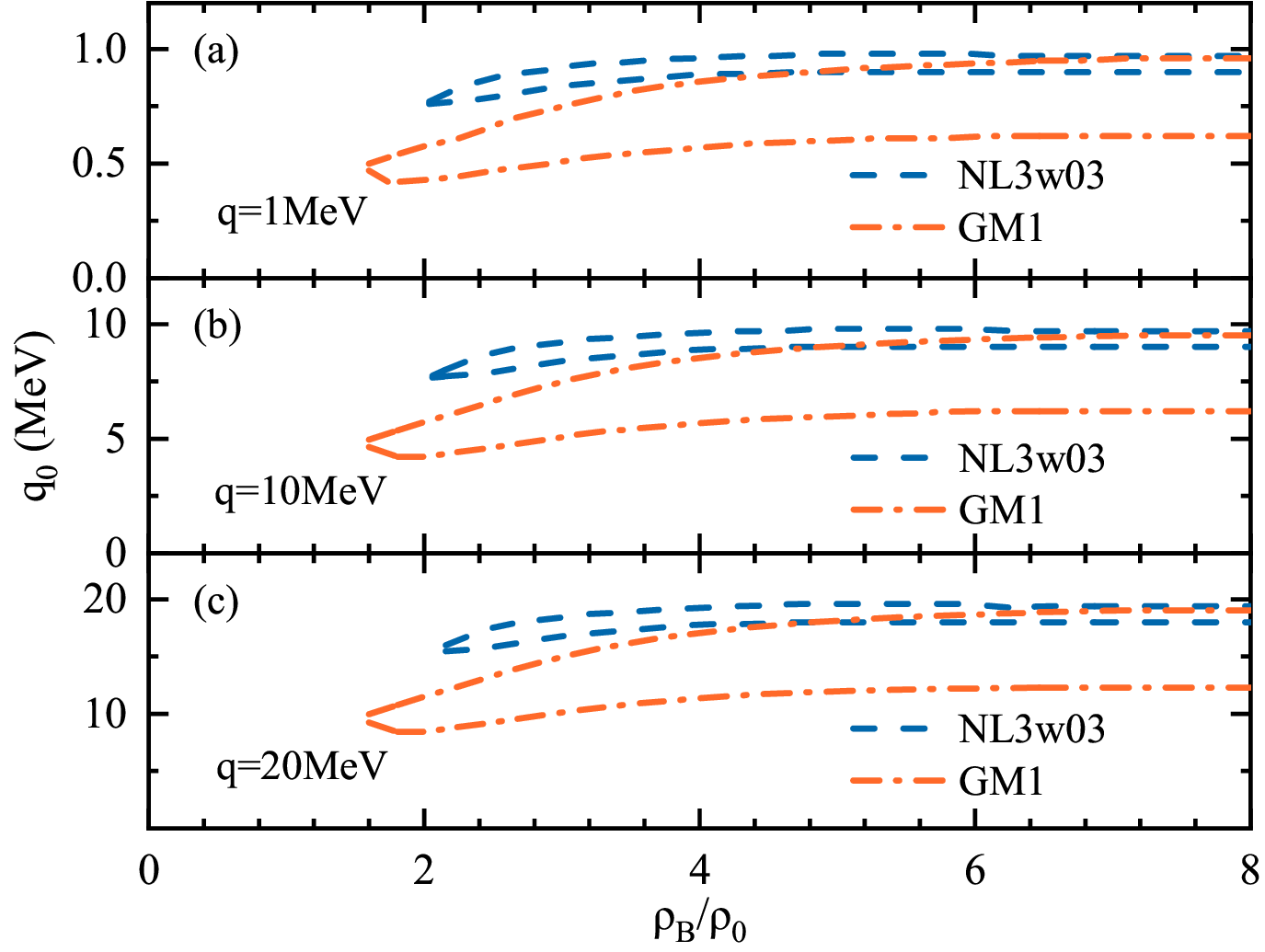}
		\centering
		\caption{\label{NLq} (Color online) High-density zero-sound modes as a function of density with the NL3w03 and GM1. (a)-(c) are the results for  momenta $q=$1, 10, and 20 MeV, respectively.}
   	\end{figure}

\begin{figure}[htbp!]
		\includegraphics[width=0.48\textwidth]{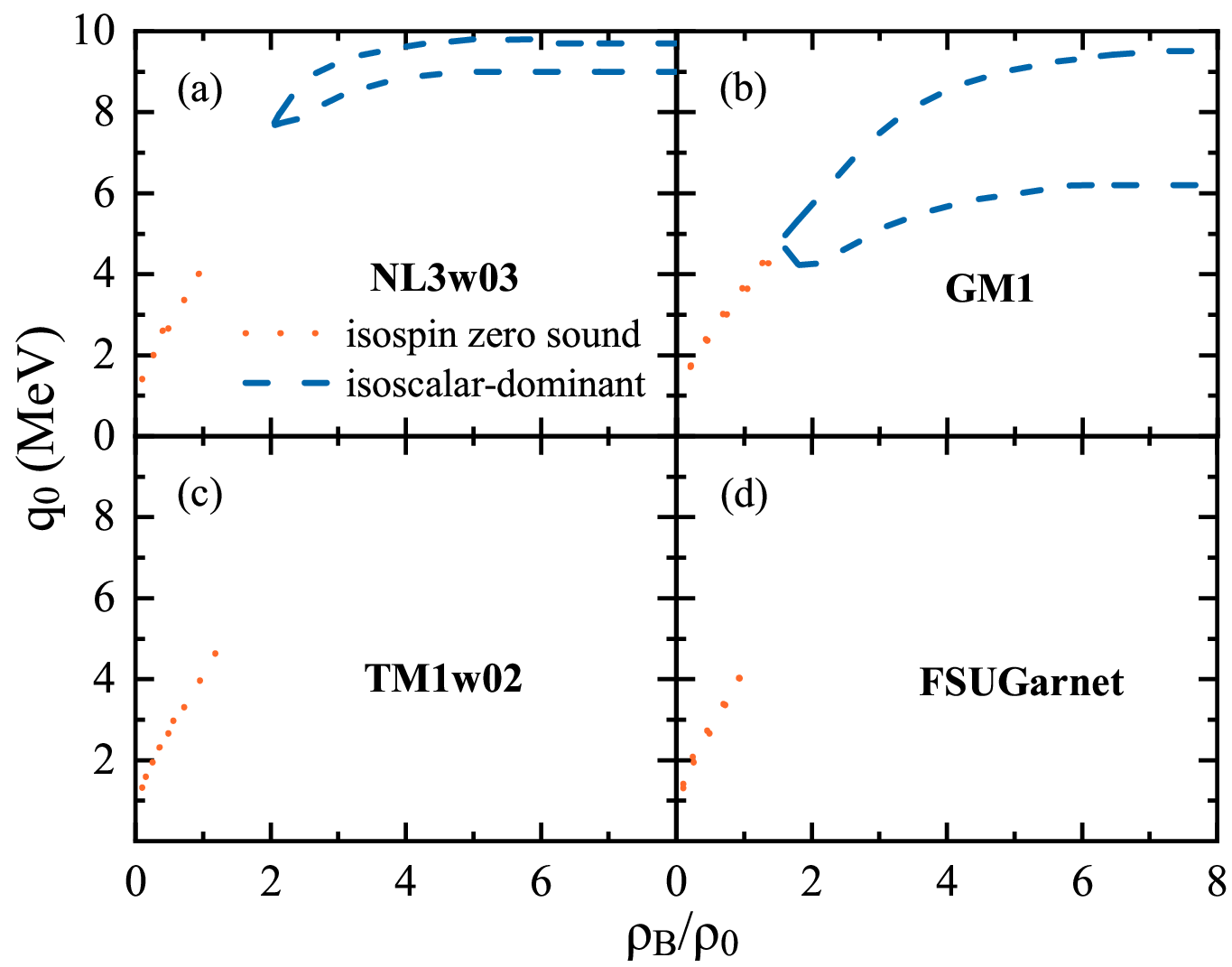}
		\centering
		\caption{\label{zs} (Color online) Zero-sound modes  ($q=10$ MeV) with various RMF models.  High-density (isoscalar-dominant) zero sounds and isospin zero sounds are distinguished by  different colors.}
	\end{figure}

	Zero sounds are usually the collective oscillation modes following  the dispersion relation at small momentum transfer $q$. We check the zero-sound modes at various small $q$=1, 10, and 20 MeV and find that the results are similar, as shown in  Fig.~\ref{NLq}. For larger momenta, the onset density of zero-sound mode increases moderately. For instance, with the model NL3w03, the appearance of zero-sound modes at the density of 2.2 $\rho_0$ is observed at $q_0=61.2$  MeV for $q=80$ MeV, in comparison to the onset density 2.05 $\rho_0$ at  $q_0=7.7$  MeV for  $q=10$ MeV. For numerical concision, the momentum $q=10$ MeV is typically chosen in the following calculation, and it is taken in the relevant  figures below, unless otherwise indicated. Represented in Fig.~\ref{zs} are the zero-sound modes with the  RMF models in symmetric nuclear matter. At low densities, it is shown that there are two close branches of the zero-sound modes in all models. Intrigued  by the $\rho$ meson primarily~\cite{matsui1981fermi}, they are labeled as the isospin zero sound. The dominance of the isovector contribution arises from   the quadratic term of the  polarization in $\varepsilon_{v}$ in Eqs.~(\ref{eqeL1}) and (\ref{eqeL2}) due to the inclusion of the $\rho$ meson exchange without which the zero point of the dielectric function does not appear at low density for the significant cancellation between the vector and scalar  potential strengths ($g_\omega^2/m_\omega^2-g_\sigma^2/m_\sigma^2$) in the presence of the scalar-vector mixing polarization $\Pi_m$~\cite{chin1977relativistic,lim1989collective}. Specifically, the longitudinal polarization in Eq.(\ref{eqeL1}) tends to have a sharp peak at low densities, in contrast to the flat distribution at high densities, and the significant magnitude of the peak ensures the sufficient cancellation of two terms in Eq.(\ref{eqeL2}) and  the appearance of the zero points.  On the contrary, a sharp difference in the zero-sound modes at high density that are dominated by the  isoscalar interactions exists for models with different stiffness. With NL3w03 at $\rho_B >2 \rho_0$ and GM1 at $\rho_B >1.6 \rho_0$, there are two zero-sound modes, which appear to be similar to those in the Walecka model~\cite{lim1989collective}, while no zero sound is found even at considerably high density in TM1w02 and FSUGarnet.  The lower curves of the zero-sound modes in NL3w03 and GM1 are always damped, abiding by $\Im\Pi\ne 0$, which allows the decay of the collective modes into real particle-hole pairs, or particle-antiparticle pairs if the vacuum polarization is involved.
	
The onset of the zero sounds is closely relevant to the stiffness of the EOS, as we see from Fig.~\ref{EOS} that the EOSs with the NL3w03 and GM1 are clearly stiffer than those with the TM1w02 and FSUGarnet.  Actually,  the zero-sound modes do necessarily need the repulsive interaction dominance~\cite{chin1977relativistic,lim1989collective}. The strong net repulsion after the partial cancellation by the scalar attraction in NL3w03 and GM1 is in support of the onset of the zero sounds  at high density, while the vector potentials with the TM1w02 and FSUGarnet are significantly weakened by the $\omega$ meson self-interaction, declining the zero sounds  at high density.

\begin{figure}[htbp!]
		\includegraphics[width=0.48\textwidth]{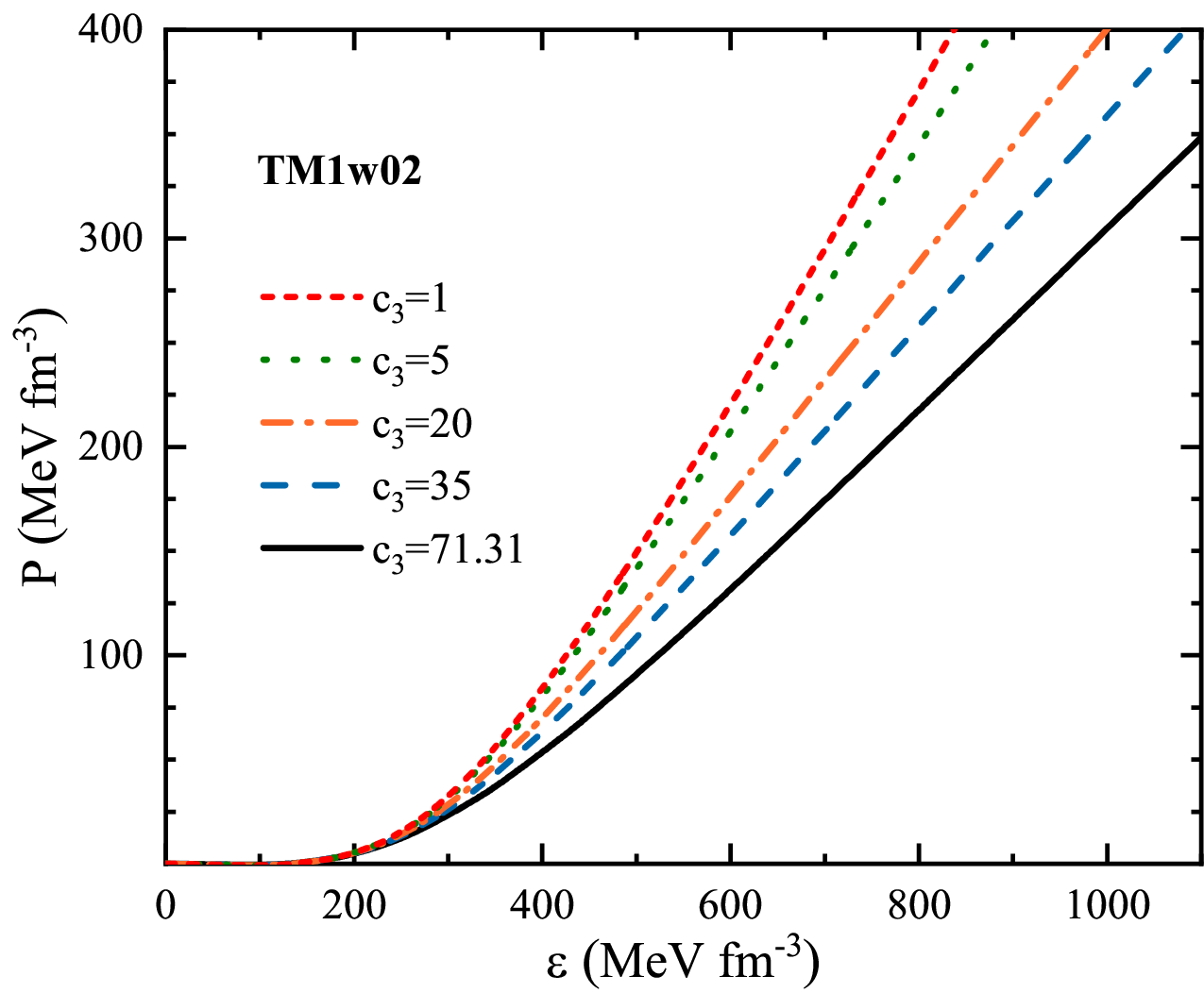}% Here is how to import EPS art
		\centering
		\caption{\label{TM CEOS} (Color online) Relation between the pressure and energy density for various EOSs with different nonlinear parameters $c_3$ in the TM1w02.}
	\end{figure}

\begin{figure}[htbp!]
		\includegraphics[width=0.48\textwidth]{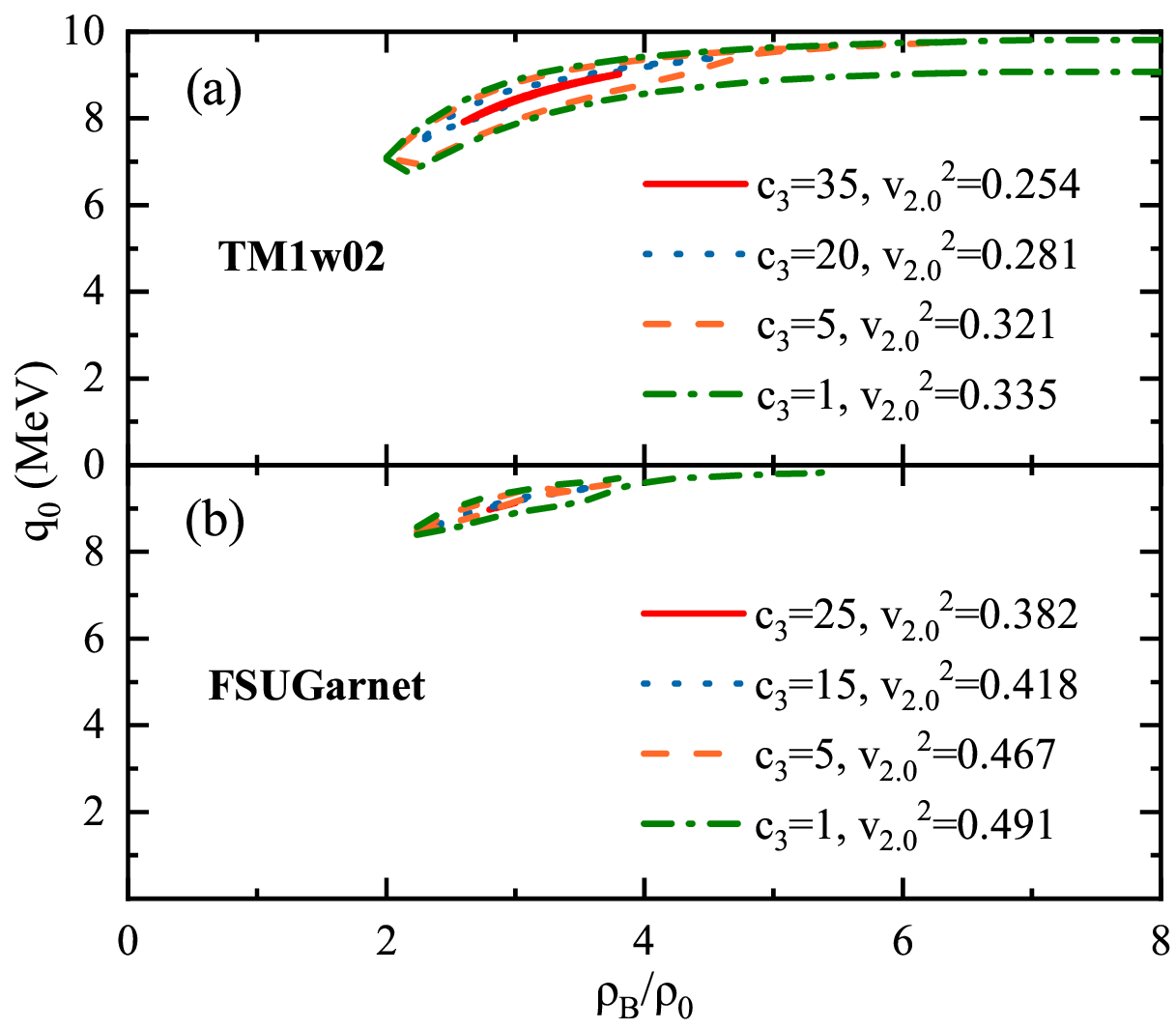}% Here is how to import EPS art
		\centering
		\caption{\label{C3 ZS} (Color online) High-density zero-sound modes with various EOSs based on the  TM1w02 and FSUGarnet.  With decreasing $c_3$ to 35 in TM1w02 and 25 in FSUGarnet, the isoscalar-dominant zero sounds appear.}
	\end{figure}

	To further clarify the relation between the stiffness of the nuclear EOS and the zero-sound modes at high density, we need to examine the zero-sound modes for various stiffness of the EOS. It is known that one can adjust the stiffness of the EOS through  the meson self-interacting terms~\cite{boguta1977relativistic,sugahara1994relativistic,todd2005neutron}, while in this work we alter the stiffness of the high-density EOS by predominantly readjusting the coupling strength of the  $\omega$ self-interacting term in Eq.(\ref{eq-L}).   In order to keep the incompressibility unchanged, other coupling constants in the model should be  moderately modified by less than 5\%~\cite{yang2017novel}. As a first example, here we carry out the model adjustment based on the original parameter sets TM1w02. The readjusted parameter sets are given in Table~\ref{tab:c3} where the sound velocities at $\rho=2.0\rho_0$ and 2.5$\rho_0$ are also presented to indicate the stiffness of EOS, and the corresponding EOSs of TM1w02 are plotted in Fig.~\ref{TM CEOS}. Since the low-density EOS is little changed, the isospin zero-sound modes just change slightly with the decrease of $c_3$.  As the EOS becomes stiffer by lowering the parameter $c_3$, we can see from the upper panel of Fig.~\ref{C3 ZS} that  the high-density  zero-sound modes start to appear at $c_3=35$ with a $v_{2.0}^2$=0.254 and gradually become expanded. The  appearance of the isoscalar-dominant zero sounds at high density is tightly associated with the nuclear effective interactions, as the reduction of the parameter $c_3$ significantly lowers the effective mass of the $\omega$ meson and strengthens the contribution from  the $\omega$ meson propagator  in the dielectric function. More specifically, the appearance of the zero-sound modes at high density requires the inequality for $\chi_\omega$: $|\chi_\omega|>(1+2\chi_\rho\Pi_L)/[2\Pi_L(1+2\chi_\rho\Pi_L)]$, according to Eq.(\ref{eqeL1}).  This inequality justifies the occurrence of the zero-sound modes for the stiffer EOS, obtained with the smaller $\omega$ effective mass by reducing the parameter $c_3$. With the FSUGarnet, the high-density zero sounds appear similarly with a stiffening of the EOS by reducing $c_3$.   As shown in the lower panel of Fig.~\ref{C3 ZS}, the high-density zero sounds appear once the parameter $c_3$ reduces to 25 or less and correspondingly the sound velocity squares $v^2_{2.0}$ increases up to 0.382, also see Table~\ref{tab:c3} for the readjusted parameters. It is worth noting that the sound velocity  corresponding to the onset of the zero sounds in FSUGarnet  is higher than that in TM1w02. This difference is mainly attributed to the negative $\sigma$ meson self-interacting coupling constant $g_3$ in FSUGarnet that is also quite different by stiffening the EOS through reducing $c_3$. The various $g_3$ of FSUGarnet in Table~\ref{tab:c3} give rise to a much reduced scalar meson effective mass $m_{\sigma}^*$ [see Eq.~\ref{eq:ms}] and affect the scalar meson propagator [see Eq.~\ref{eq:Xs}[.

	\begin{table*}
		\caption{\label{tab:c3}The parameters for various nuclear EOS’s based on TM1w02 and FSUGarnet. The unlisted parameters are the same as those of the corresponding parameter set. The sound velocity square $v_{2.0}^2$ at $2\rho_0$ and $v_{2.5}^{2}$ at $2.5\rho_0$ are listed.  Also given are the maximum mass of neutron stars and radius of the $1.4M_\odot$ star with the composition
of neutrons, protons, and electrons.
		}
		\begin{ruledtabular}
			\begin{tabular}{cccccccccc}
				
				Models&$c_{3}$&$g_{2}$&$g_{3}$ &$g_{\sigma}$&$g_{\omega}$&$v_{2.0}^{2}$&$v_{2.5}^{2}$&$M_{max}(M_{\odot})$&$R_{1.4}$(km)\\ \hline
				
				&71.31&7.233 &0.618  &10.029 &12.614 &0.213 &0.313 &2.120&13.487 \\
				&35& 8.904 & -11.587  &9.984 &12.451 &0.254 &0.370 &2.292&13.647\\
				TM1w02&20& 9.609 & -16.850  &9.961 &12.380 &0.281 &0.419&2.401&13.719 \\
				&5& 10.338 &  -22.266  & 9.938 & 12.307 &0.321 &0.494 &2.558&13.797\\
				&1& 10.537 &  -23.738  & 9.932 & 12.287 &0.335 &0.523 &2.617&13.818\\
                &137.98&9.576&-7.207&10.505&13.700&0.228&0.314&2.066&11.706\\
                &25&11.089&-28.575&10.174&13.084&0.382&0.512&2.494&13.294\\
                FSUGarnet&15&11.252&-30.631&10.141&13.022&0.418&0.560&2.581&13.337\\
                &5&11.420&-32.720&10.107&12.958&0.467&0.627&2.695&13.382\\
                &1&11.489&-33.566&10.093&12.932&0.491&0.663&2.752&13.401\\
			\end{tabular}
		\end{ruledtabular}
	\end{table*}	
	
\begin{figure}[htbp!]
		\includegraphics[width=0.48\textwidth]{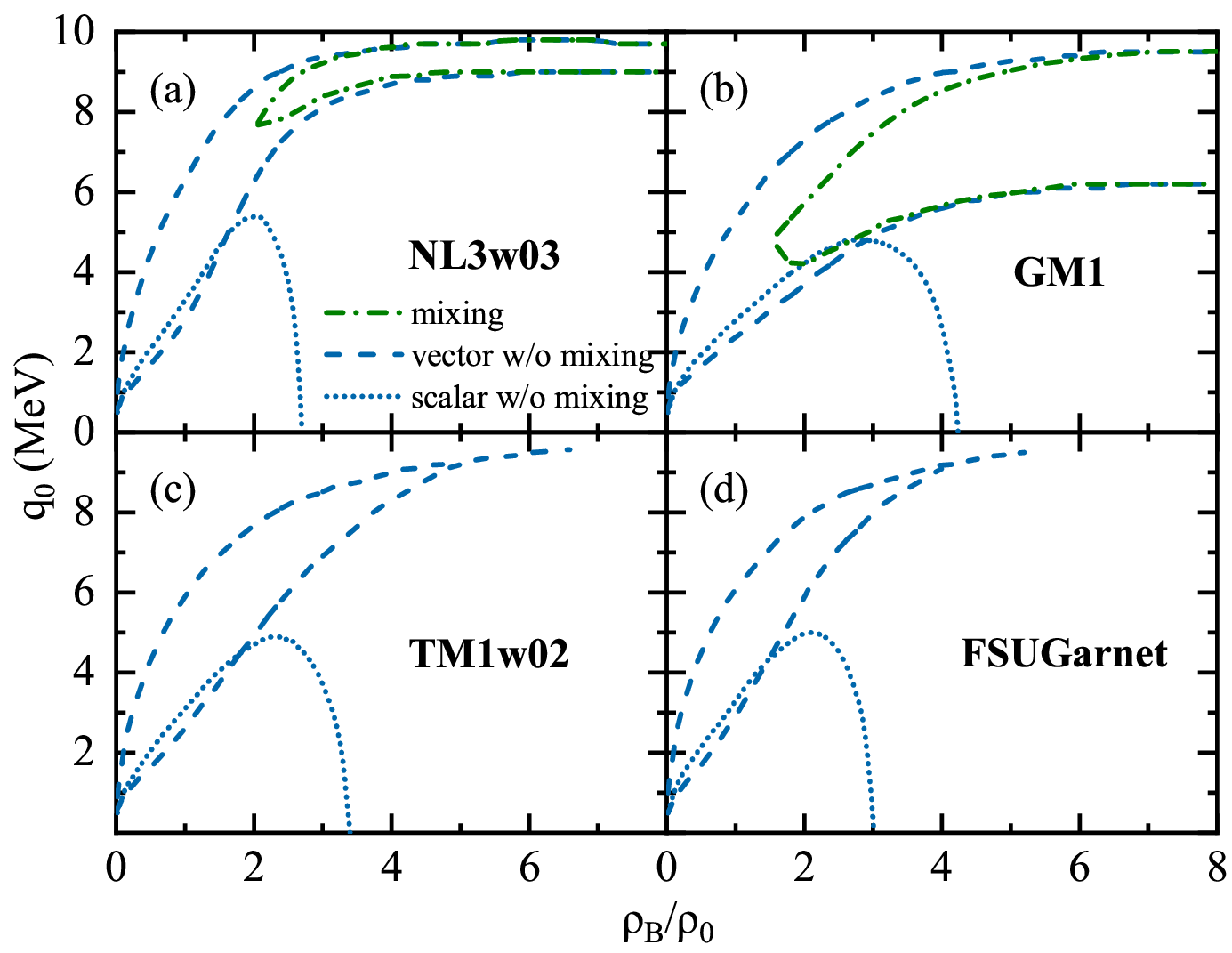}% Here is how to import EPS art
		\centering
		\caption{\label{NL msv} (Color online) Isoscalar-dominant zero-sound mode with the NL3w03, GM1, TM1w02, and FSUGarnet. Dot-dashed curves are with scalar-vector mixing, the dashed curve represents zeros from the vector meson branch ($\varepsilon_v$), and the dotted curve shows the zeros from the scalar meson propagation ($\varepsilon_s$).}
	\end{figure}

It is worth mentioning the role of the scalar-vector mixing polarization ($\Pi_m$) in the onset of the zero sounds. Without the $\Pi_m$, zero sounds arising from the poles of the vector meson branch in $\varepsilon_v$ of Eq.(\ref{eqeL2}) appear even at low densities with all of the parameter sets NL3w03, GM1, TM1w02, and FSUGarnet, as shown in Fig.~\ref{NL msv}. The dotted curve shows the poles from the scalar meson which are related to the instability mode~\cite{lim1989collective}. Note that, for FSUGarnet, there are no poles from the vector meson at high density due to the weak repulsive potential. In comparison to the results in Fig.~\ref{zs}, we can conclude that the $\Pi_m$ plays a necessary role in the cancellation between the contributions from the vector and scalar interactions,  thereby influencing the onset of the zero sounds.	

	\begin{figure}[htbp!]
		\includegraphics[width=0.48\textwidth]{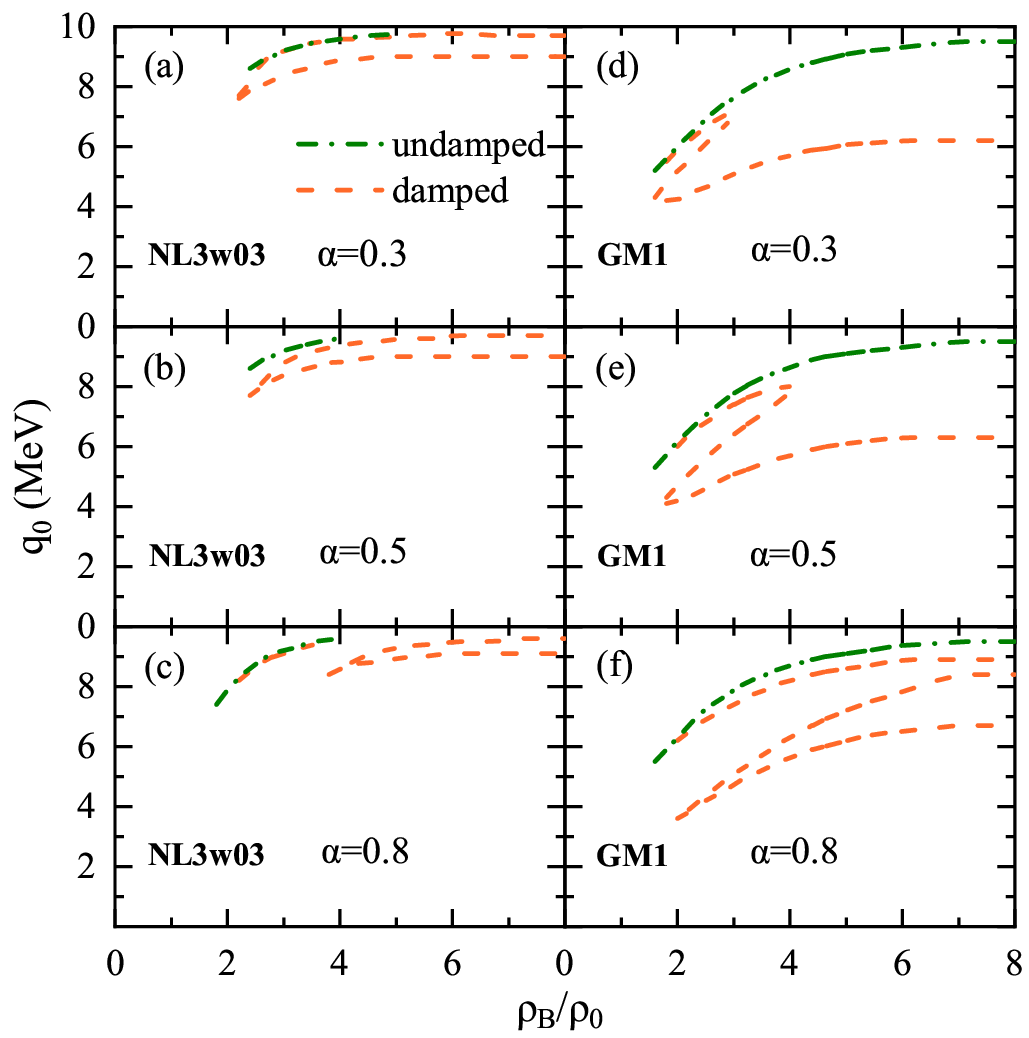}% Here is how to import EPS art
		\centering
		\caption{\label{NL3 ASY} (Color online) Zero-sound modes with various asymmetry parameters $\alpha$ in the NL3w03 and GM1. Damped and undamped zero-sound modes are distinguished by different colors and line types.}
	\end{figure}
		
		\begin{figure}[htbp!]
		\includegraphics[width=0.48\textwidth]{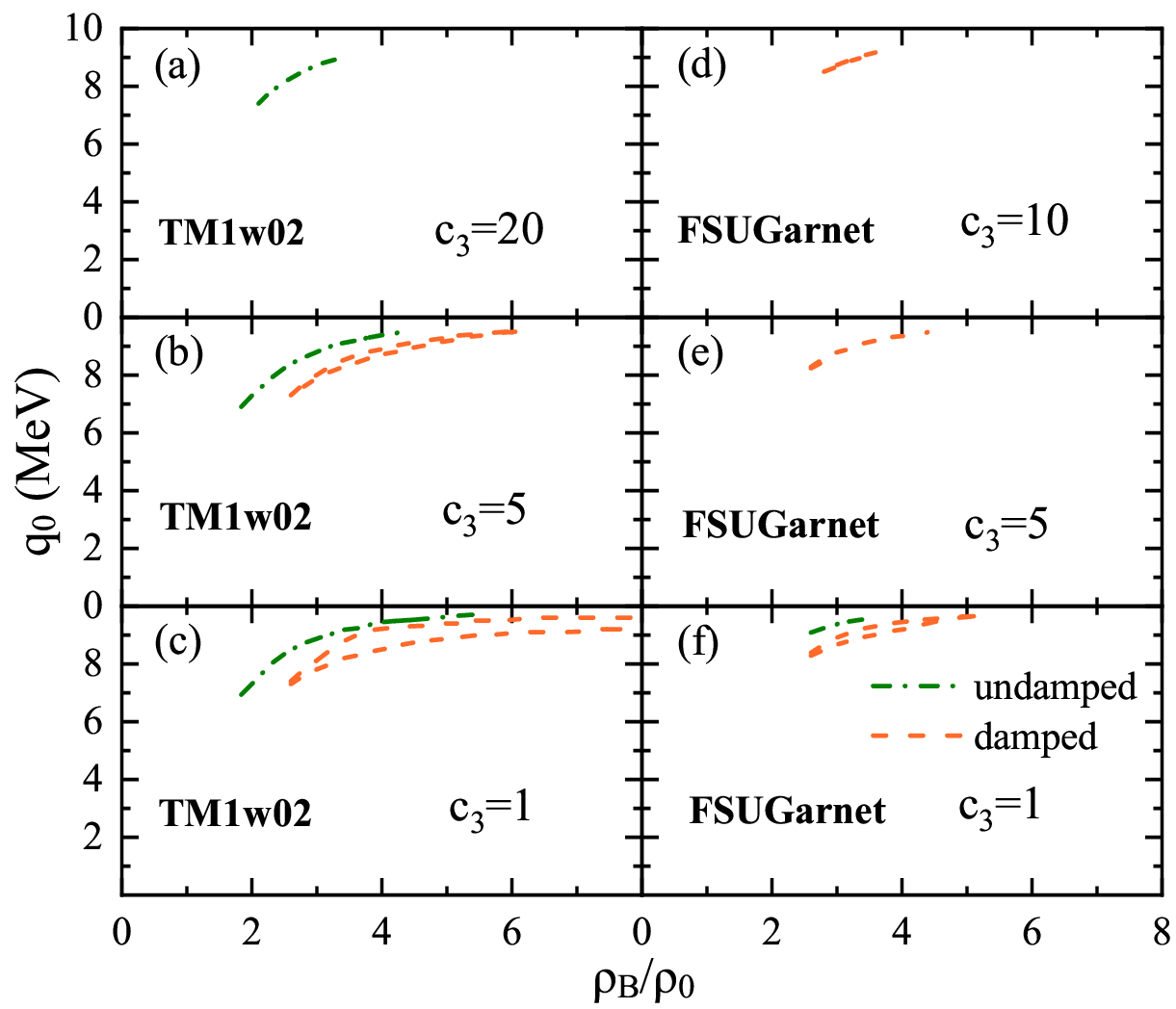}
		\centering
		\caption{\label{TM1 ASY} (Color online) Zero-sound modes in asymmetric nuclear matter ($\alpha=0.5$) with the readjusted parameter $c_3$ in the TM1w02 and FSUGarnet.}
	\end{figure}

Above is the analysis in symmetric matter, and now we turn to the results in asymmetric matter. Since the RMF approximation almost gives rise to the same nucleon as that in symmetric matter, the main difference of the dielectric function  arises from the separate proton and neutron Fermi momenta in the integrations of the polarizations and from the isospin asymmetry dependence of the  meson effective masses due to the isoscalar-isovector coupling. As a result, the zero-sound modes in asymmetric matter, obtained from the zeros of the dielectric function, can be more than those in symmetric matter, as shown in Fig.~\ref{NL3 ASY}.  Figure~\ref{NL3 ASY} shows that the density profiles of the isoscalar-dominant zero sounds with the stiff models NL3w03 and GM1 are quite different at various isospin asymmetries [$\alpha=(\rho_n-\rho_p)/\rho_B$]. This difference arises  primarily from the diverse values of $m_{\sigma}^*$ and correspondingly different scalar meson propagators in two models.  Meanwhile,  no zero sounds are found in soft models TM1w02 and FSUGarnet  in asymmetric matter. Accordingly, in asymmetric matter the soft and stiff EOSs can be similarly distinguished  with the onset of the zero-sound modes.     In models with the soft EOS, the occurrence of zero-sound modes  in asymmetric matter needs the stiffening of the EOS, similar to that in symmetric matter.   For instance, at $\alpha=0.5$,  the  zero-sound modes in TM1w02 (FSUGarnet) start to appear, as shown in  Fig.~\ref{TM1 ASY}, with stiffening the  EOS by reducing the value of  the parameter $c_3$ to 20 (10).
On the other hand, the isospin zero-sound modes vanish gradually with increasing $\alpha$, independent of EOS stiffness. We will leave a detailed check for the isospin zero sounds in a separate work, since it is a little digressive of the theme of this work concerning the sensitivity to the  EOS stiffness.

The dependence of the zero sound occurrence on the stiffness of the EOS in dense asymmetric matter has experimental implications to possible observables concerning the heavy ion collisions.   As the centroid energy of zero sounds is comparable with the temperature or some thermalized energy of the colliding system, the formation of dense matter, the subsequent thermalization, and the late relaxation period with the particle emissions, evaporations and fragmentation ought to be affected by the zero-sound modes and may have signals from the emitted particles~\cite{sorensen2023dense}. An intensive investigation of the heavy  ion collision and detailed comparison   with the experimental signals can hopefully be used for detecting the zero sound occurrence and its relation to the stiffness of the EOS. In addition, it is interesting to note that zero-sound modes are important in the low-energy fusion reactions near threshold. At astrophysical energies, various zero-sound modes can strikingly enhance the subthreshold fusion rate in the astrophysical nucleosynthesis~\cite{tumino2018increase,zhang2022measurement}. In the hot-fusion reactions for the synthesis of superheavy nuclei, it also shows some evidence that the zero-sound modes might take a part in increasing the survival rate of the compound nucleus prior to the fission\cite{zhu2016thermal}.

		\begin{figure}[htbp!]
	\includegraphics[width=0.48\textwidth]{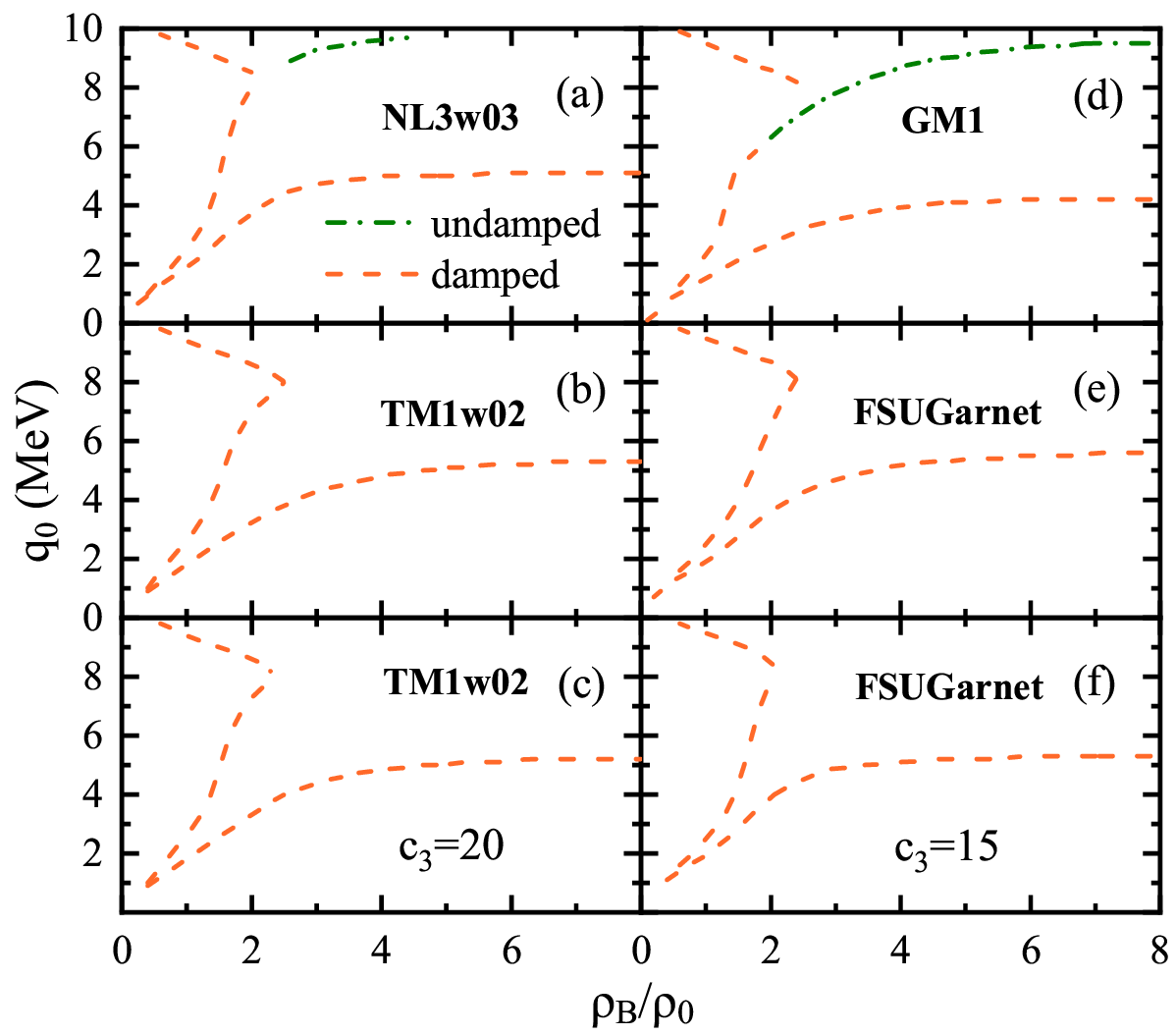}% Here is how to import EPS art
		\centering
		\caption{\label{neutron star} (Color online) Zero-sound modes with various RMF models in neutron star matter. Presented in the lowest panels (c) and (f) are zero-sound modes in the stiffened TM1w02 with $c_3=20$ and FSUGarnet with $c_3=15$, respectively.}
	\end{figure}

Below, we discuss the behavior of zero-sound modes in neutron star matter with the simple composition of neutrons, protons, and electrons at $\beta$ equilibrium. Zero-sound modes in neutron stars can possibly be associated with neutron star cooling~\cite{bedaque2003goldstone,Aguil2009prl,leinson2011zero} and pulsar radiation~\cite{svidzinsky2003radiation}, their correlation with the sound velocity can directly point to the stiffness of the EOS~\cite{tews2018constraining,bedaque2015sound}. Therefore, investigating the connection between zero-sound modes and the EOS of  neutron star matter during the multimessenger era is of significance  for our understanding of relevant astrophysical phenomena. Shown in Fig.~\ref{neutron star} are the zero-sound modes in neutron star matter with various soft and stiff models.  As shown in  Fig.~\ref{neutron star}, damped zero-sound modes are similar  with various RMF models, almost independent of the stiffness of the EOS, while undamped modes exhibit diverse results. Undamped zero-sound modes exist only in the stiff models NL3w03 and GM1. Even the models with the soft EOS are made much stiffer by reducing $c_3$, undamped zero-sound modes still do not come up. Due to the complexity of neutron stars, including  the variations in the  isospin-dependent interactions, isospin asymmetry, and internal compositions with different models, the correlation between zero-sound modes and the stiffness of the EOS is weakened. Numerically, the weakening of the correlation is not surprising, since the subtle cancellation for the zeros in the dielectric function in asymmetric matter,  see Eqs.~(\ref{eqeL1}) and (\ref{eqdi1}), is violated by including the electrons and photon exchange in neutron star matter at high isospin asymmetry.
 
	\section{\label{sec:level4}SUMMARY}
	In this work, we have investigated in the RMF theory the properties of zero-sound modes in nuclear matter  through the response function in the relativistic RPA. The nuclear models we choose are roughly classified into two categories of the soft and stiff EOSs, and the readjustment of the meson self-interaction couplings is conducted to modulate the stiffness of EOS. It shows that in symmetric nuclear matter  all selected RMF models produce the isospin zero-sound modes at low density similarly, and  the  isoscalar-dominant zero sound is absent at low density.  It is found that  sharp difference exists in the zero-sound modes at high density due to different stiffness of the EOS. The zero sounds at high density appear for the stiff EOS and in contrast are absent for the soft EOS. As the density increases, the contribution of the $\rho$ meson to the  zero-sound modes diminishes, while the interaction of the $\omega$ meson develops gradually to be dominant. At high density, the isoscalar-dominant zero sounds arise in the stiff models NL3w03 and GM1 with their repulsive  $\omega$ field being linear in density. As the repulsion is weakened  by the nonlinear self-interaction of the $\omega$ meson, the zero sounds disappear at high density, as typically manifested by the models  TM1w02 and FSUGarnet that are characteristic of the softening of the repulsion. In these soft models, the high-density zero sound reappears when the EOS stiffens by reducing the nonlinear self-interaction of the $\omega$ meson appropriately.  Apparently, whether the isoscalar-dominant zero sounds occur at high density or not can  serve as a significant probe to categorize the stiffness of the high-density EOS which suffers the large uncertainty. It also indicates that the density spread of the isoscalar-dominant  zero sounds at high density is sensitive to the stiffness of EOS. In addition, the implications and effects of zero sounds are also discussed in heavy ion collisions and neutron stars. In particular, for the  zero-sound mode with the centroid energy being comparable with the temperature of compressed matter,  detecting the zero sound occurrence and its relation to the stiffness of the EOS can be hopeful through  intensive investigation of the heavy  ion collision and detailed comparison  with the experimental signals.
	
	\section*{ACKNOWLEDGMENTS}
We thank Drs. Rong-Yao Yang, Si-Na Wei, Profs. Gao-Chan Yong, and Zhao-Qing Feng for useful discussions. This work was supported in part by the National Natural Science Foundation of China under Grant Nos. 11775049 and 12375112. The Big Data Computing Center of Southeast University is acknowledged for providing the facility support on the partial numerical calculations of this work.

	\bibliography{ref}

\end{document}